\documentclass[12pt]{article}

\usepackage[top=1in, bottom=1in, left=1in, right=1in]{geometry}

\usepackage{graphicx}

\usepackage[space]{grffile}

\usepackage{gensymb}

\usepackage{textcomp}

\usepackage{times}

\usepackage{float}

\usepackage{setspace}

\usepackage[super]{natbib}

\usepackage{ragged2e}

\usepackage{microtype}

\usepackage{breakcites}

\usepackage{url}

\usepackage{makecell}

\usepackage{lineno}

\usepackage{abstract}

\makeatletter
\newcommand{\mypm}{\mathbin{\mathpalette\@mypm\relax}}
\newcommand{\@mypm}[2]{\ooalign{%
  \raisebox{.1\height}{$#1+$}\cr
  \smash{\raisebox{-.6\height}{$#1-$}}\cr}}
\makeatother

\title{Towards the development of a macro-structured water-repellent surface}
\author{ Regulagadda Kartik$^{\dagger}$, Shamit Bakshi$^{\dagger *}$, Sarit Kumar Das$^{\dagger \ddagger}$\\
$\dagger$ Department of Mechanical Engineering, Indian Institute of Technology,\\ Madras, India.\\
$\ddagger$ Current Address: Department of Mechanical Engineering, Indian Institute of\\ Technology, Ropar, India. \\
Email: kartik25192@gmail.com\\
Email: shamit@iitm.ac.in\\
Email: skdas@iitrpr.ac.in\\
Phone: +91-44-22574700}
\date{}

\begin{document}

\maketitle

\begin{abstract}
Drop-surface interaction is predominant in nature as well as in many industrial applications. Freezing rain is the frequent origin of ice accretion on surfaces. Superhydrophobic surfaces show potential for anti-icing applications as they exhibit complete drop rebounce. Nonetheless, drop shedding has to take place before freezing for effective functioning. Recently, introducing a macro-ridge to break the hydrodynamic symmetry, has been shown to reduce the residence time on the surface of a bouncing drop. However, for a practical application the surface must be decorated with a series of ridges so that most of the drops actually encounter the ridges and lift-off rapidly. Here we show that a parallel neighbor ridge can influence the dynamics of recoiling. Ridge spacing plays a key role on the performance of surface to reduce the residence time. This finding can be of great significance for the development of macro-ridged anti-icing surfaces.
\end{abstract}

Superhydrophobic surfaces have gained significant attention over the last two decades for astounding applications like self-cleaning \citep{Quere2008, Barthlott1997, Marmur2004}, anti-icing\cite{Cao2009, Jung2011, Liu2014, Bird2013, Liu2016, Gauthier2015}, drop-wise condensation, low-friction flows etc. Today's emerging technologies have enabled the surface morphology manipulation at micro and nano-scale and thereby fabricating surfaces with desired roughness and wettability \cite{Li2007}. The complex momentum and energy exchange between the drop and surface governs the hydrodynamics of impact. The residence time \cite{Richard2002} and spread behavior \cite{CLANET2004} on the surface are studied to capture the complete bouncing effect. The phenomenon involves very high velocity and acceleration.

The spectacular properties offered by superhydrophobic surfaces are often limited by the robustness of surface under harsh environmental conditions \cite{Jung2012, Bartolo2006, Maitra2014, Varanasi2010, Meuler2010, Kulinich2009}. Macroscale features of surface \cite{ Liu2014, Bird2013, Liu2015a, Vollmer2014, Weisensee2016} are found to significantly alter the hydrodynamics of impinging drops. Decorating the surface with a macro-ridge breaks the hydrodynamic symmetry during the recoiling of drop. The residence time is found to reduce upto 45\% with a step-like variation in velocity \cite{Gauthier2015} which is in contrast with macroscopically flat surfaces. However, centering of impacts is essential for residence time reduction. Here we show that introducing parallel ridges in close proximity to each other will "suffocate" the drop lift-off and on the other hand a judicious selection of the nearness of the neighboring ridge will help to retain its benefit. The peak-to-peak distance $p$ is varied systematically and this additional length scale is shown to govern the ability to rapidly shed water drops.
 
\subsection*{Results}
\subsubsection*{Hydrodynamics} Five polished aluminium substrates with different $p$ are manufactured. An isosceles wedge shaped ridge is machined with cross-sectional base width of 1 mm and height of 0.5 mm. A characteristic parameter $k_n$ is defined as, $k_n = D_0/p$ where $D_0$ is the initial diameter of the drop and $p$ is $D_0/n$. The selected values of $n$ are 0, 1, 2, 3 and $\infty$.  Note that $k_0$ and $k_\infty$ represents single ridge and flat subsrates respectively. All the substrates are coated with commercially available superhydrophobic coating Ultra Ever Dry from UltraTech International, Inc. The advancing and receding contact angles observed are $169\degree \mypm 2\degree$ and $166\degree \mypm 2\degree$ respectively and roll-off angles were found to be $< 5\degree$ which signify the fakir state \cite{Lafuma2003}.

\begin{figure}[H]
	\begin{center}
  \includegraphics[width = 16cm, height = 9cm]{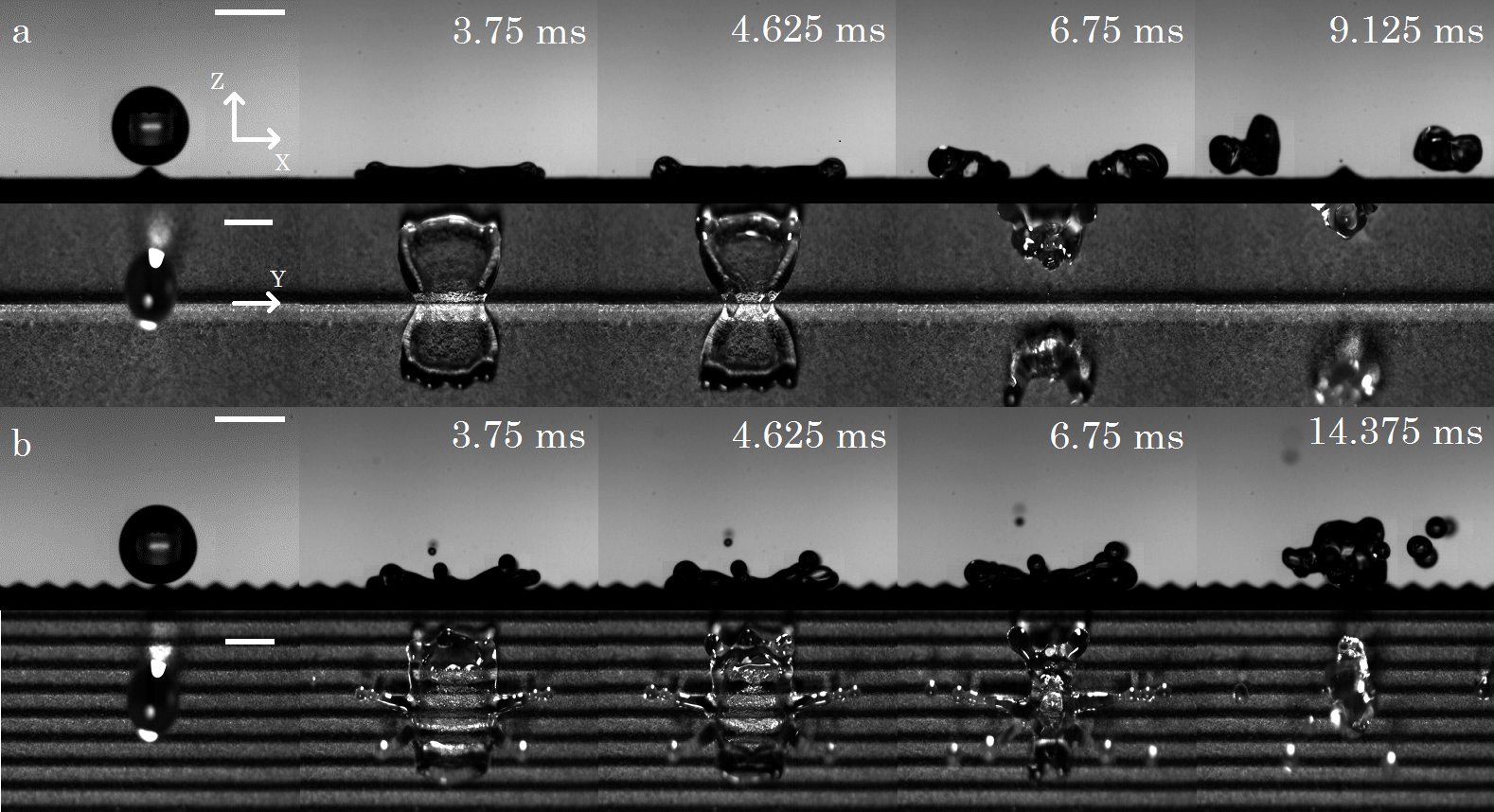}
 \caption{Side and top views for impact on $k_0$ (a) and $k_3$ (b) substrates at $We =$ 43.8. (\textbf{a}) The liquid on the either side of ridge appear as butterfly wings at $t$ = 3.75 ms and lift-off takes place at time $t$ = 9.125 ms. (\textbf{b}) The hydrodynamics of the impact is greatly effected with the influence of neighbor ridges and Rayleigh jet formation takes place immediately after the impact. The residence time increases as bouncing is delayed. Scale bar in the inset represents 2 mm.}
  \label{overall_p_and_v}
  \end{center}
\end{figure}

Milli-Q water drops with density $\rho = 1000$ kg/m$^3$ and surface tension $\sigma = 0.073$ N/m are dispensed from a calibrated needle. $D_0$ is found to be 2.95 $\mypm$ 0.03 mm for the entire set of experiments. The height of the needle is varied to change the impact velocity $U$. The experimental range of Weber number considered is 0.4 $< We <$ 75 where $We = \rho U^2 D_0/\sigma$. The impact is captured from the side (Photron SA4) and top (Photron UX Mini 100) with a typical frame rate of 8000 fps. Image sequences of Fig. \ref{overall_p_and_v} (a and b) show the side and top views of impact for $k_0$ and $k_3$  substrates respectively for $We = $ 43.8 (details of impingement on all substrates at the same $We$ are included in the supplementary videos 1-10).

\begin{figure}[H]
	\begin{center}
  \includegraphics[scale = 0.3]{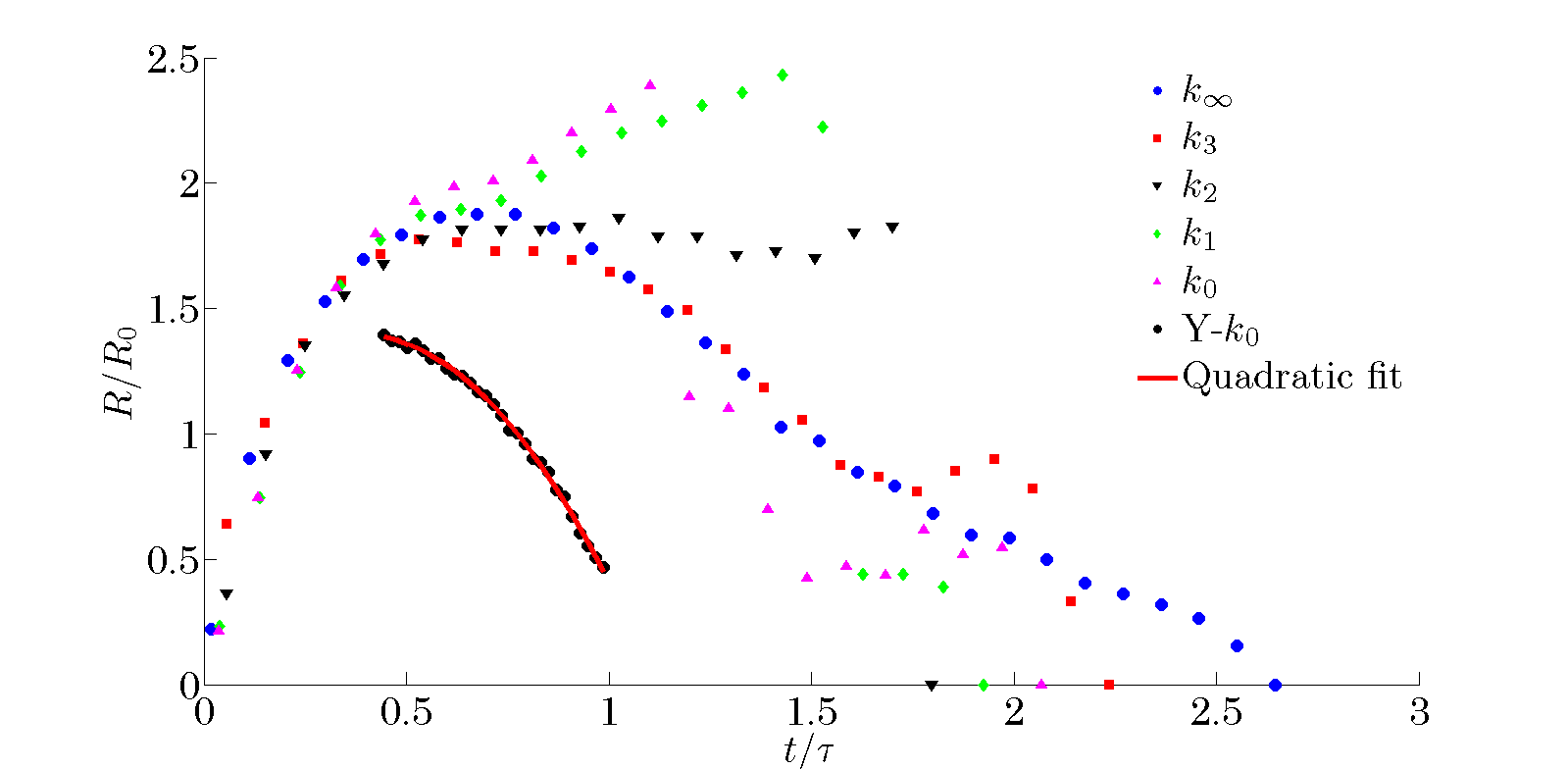}
  \caption{Radial spread in X-direction at $We =$ 25.8 for all the substrates. The black circles represent the spread length along the ridge (Y-direction) during the recoiling stage for $k_0$ substrate. The solid line represents the quadratic curve fit of the data.}
  \label{We25}
  \end{center}
\end{figure}

It can be observed that many liquid ligaments are formed during impact on $k_3$ substrate in the form of Rayleigh jet. This makes the phenomenon slightly different. The transient radial spread in X-direction of the impact for $We$ = 25.8 is shown in Fig. \ref{We25} where the direction perpendicular to the ridge is considered as X,  parallel to the ridge is considered as Y and vertically upwards perpendicular to X and Y is considered to be Z as shown in Fig. \ref{overall_p_and_v}. The inertio-capillary time scale $\tau$ in Fig. \ref{We25} is defined as $\sqrt{\rho {R_0}^3 / \sigma}$ and $R$ is the radial spread of the drop at a time instant $t$ and $R_0 = D_0/2$. The maximum radial spread ($R_{max}$) followed the scaling $R_{max}/R_0$ {\raise.17ex\hbox{$\scriptstyle\mathtt{\sim}$}}\space $We^{0.25}$ with a prefactor of 0.85 (see supplementary Fig. 1) for $1 < We < 75$ on the flat substrate which is in good agreement with Clanet et al., 2004 \cite{CLANET2004}. It can be clearly observed that, until $t/ \tau \approx 0.6$: the radial spread in X for the inertial regime of impact is not influenced by the presence of macro-ridge. Furthermore, after the end of inertial regime, the radial spread increases for $k_0$, $k_1$ and $k_2$ substrates as the liquid pumping takes place towards the "wings" from the ridge. Note that $k_3$ substrate follows similar trend as $k_\infty$ substrate until $t/ \tau \approx 1.8$. When the $We$ is further increased, film rupture can be observed even before complete ridge de-wetting (see Fig. 1a and supplementary video 10). It can also be noted that there is early and rapid retraction of the film along the ridge (Y-direction).    

\subsubsection*{Acceleration along the ridge} 
\begin{figure}[H]
\begin{center}
  \includegraphics[scale = 0.3]{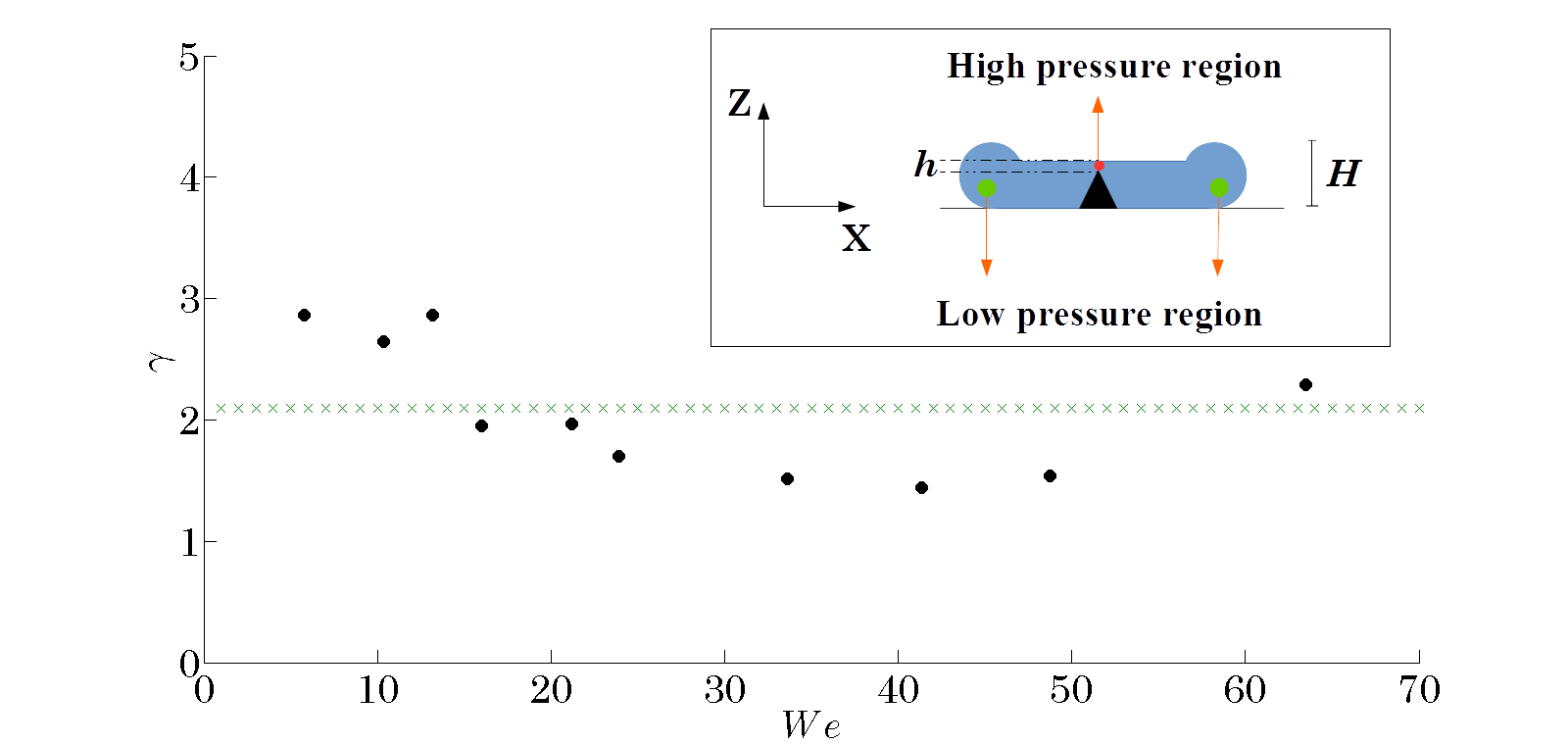}
  \caption{Acceleration ratio $\gamma$ with $We$ for $k_0$ substrate. The dashed line represents the average value of $\gamma$. Inset represents the high and low pressure regions for $k_0$ substrate and the film thicknesses $h$ and $H$.}
  \label{acceleration}
  \end{center}
\end{figure}

Gauthier et al., 2015,\cite{Gauthier2015} explained that the recoiling of liquid along the ridge has a constant acceleration kinematics which is in contrast with the Taylor-Culick velocity \cite{Taylor1959, Culick1960} observed during recoiling on a flat surface. Bird et al., 2013,\cite{Bird2013} comprehended the film thickness reduction at the ridge as a reason for rapid recoiling rates. Building upon this, we argue that the liquid mass along the ridge is pumped to the sides because of Laplace pressure increase at the ridge tip. As the liquid from all sides tries to recoil, the liquid from ridge feeds the boundary rim as the pressure gradients are favorable relative to the central film on either side of the ridge. The theoretical acceleration can be scaled as $a_{th}$ {\raise.17ex\hbox{$\scriptstyle\mathtt{\sim}$}} \space $(\sigma/h - \sigma/H)/ \rho R_0$ where $h$ is the film thickness (in Z-direction) at maximum spread along the ridge and $H$ is the maximum film thickness (in Z-direction) at maximum spread on the "wing" region observed in the experiments (see inset in Fig. \ref{acceleration}). $R_0$ is used as a length scale for the maximum spread as the "wing" region experiences the pressure difference in the X-direction. The experimental acceleration $a_{exp}$ is obtained from a quadratic fit between the contact line position and time along the ridge as shown in Fig. \ref{We25}.

For $We < 5$, the residence time reduction cannot be realized due to the low asymmetry generated during the crash in inertial regime. Hence, $a_{th}$ and $a_{exp}$ are expressed for impacts with $We > 5$. A parameter namely, acceleration ratio $\gamma$ is defined as the ratio of $a_{exp}$ and $a_{th}$, i.e. $\gamma = a_{exp}/a_{th}$. Note that $\gamma$ itself is the scale factor whose order of magnitude is found to be 2.1 $\mypm$ 0.5, which signifies the fact that the Laplace pressure gradient is the reason for ridge de-wetting. Fig. \ref{acceleration} shows the plot with $\gamma$ and $We$. Any deviations from the experimental values can be attributed to the inherent assumptions in the model. First, the film thickness at the maximum spread at time $t$ varies along the angular direction around the axis of impact. This means that the pressure gradient varies with distance from the ridge in the "wing" region. The model accounts only for the pressure gradients corresponding to the minimum and maximum film thickness. Second, the length scale under which the pressure gradient is involved, i.e. maximum spread varies with impact velocity $U$, yet the model incorporates a constant value. Third, the contact angle and the secondary radius of curvature are neglected. The receding contact angle is close to 180$\degree$ with low hysteresis and the respective secondary radius of curvature is much greater than $h$ and $H$ during most of the recoiling stage. Last, the film thickness $H$ varies with time $t$ during recoiling while $h$ remains fairly constant. Nevertheless, considering the simplicity of the model, a reasonable match with the experimental data is achieved.

The magnitude of $a_{exp}$ is found to be {\raise.17ex\hbox{$\scriptstyle\mathtt{\sim}$}} 400 m/s$^2$ for $5 < We < 30$ and {\raise.17ex\hbox{$\scriptstyle\mathtt{\sim}$}} 1000 m/s$^2$ for $30 < We < 65$. These values are much higher than the ones reported in literature which can be attributed to the ridge height deployed. In spite of this, the maximum residence time reduction is still observed to be 45\%. Thus, acceleration along ridge is governed by the ridge height while the residence time does not vary much which again emphasizes the inertio-capillary nature of the phenomenon. Note that the absolute values of acceleration may not quantify the residence time. However, the relative incremental steps in acceleration can be realized for a particular macro-ridge. Furthermore, the macro-texture height should be sufficient enough to generate hydrodynamic asymmetry. 

\subsection*{Discussion}
\begin{figure}[H]
\begin{center}
  \includegraphics[scale = 0.3]{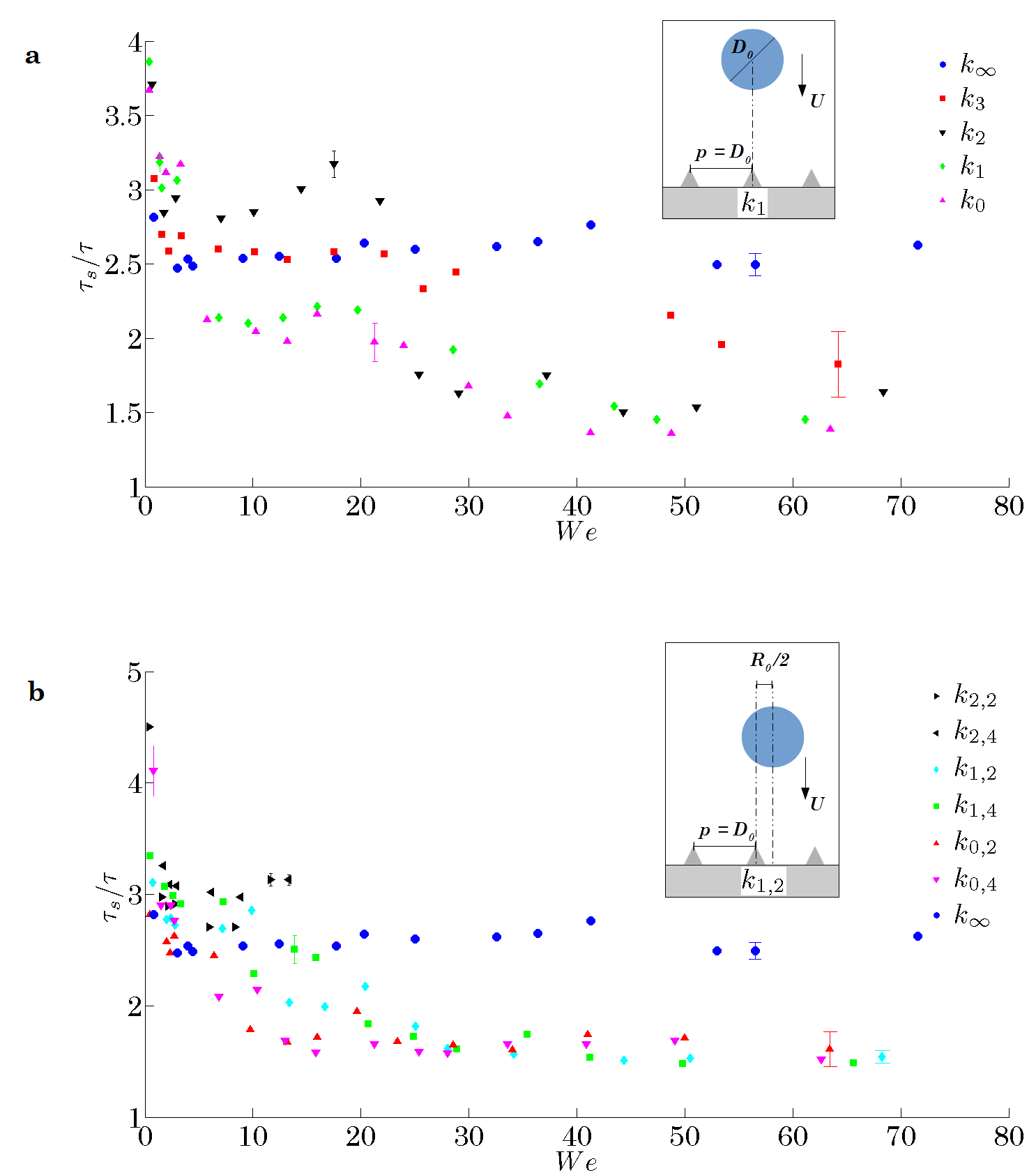}
  \caption{Plot showing the non-dimensionalized residence time with $We$ for impacts without and with offsets (see inset). (\textbf{a}) $\tau_s$ represents the residence time on the substrate. It can be observed that the residence time is always higher for $k_3$ substrate compared with $k_0$ and $k_1$ substrates. For $5 < We < 25$, the residence time on $k_2$ is higher than $k_{\infty}$ and for $We > 25$, it follows same trend as $k_0$ and $k_1$ substrates. (\textbf{b}) The offset distances considered are $R_0/2$ and $R_0/4$ from the peak of the ridge. The notation $k_{n,m}$ represents the impact on substrate $k_n$ with offset $R_0/m$. Residence time of $k_{\infty}$ substrate is included as a reference. For $k_{2,2}$ and $k_{2,4}$ the residence time is higher even when compared to $k_{\infty}$ substrate at $We < 14$ . Error bars represent the uncertainty of $\tau_s/ \tau$.}
  \label{contacttime_with_and_without_offset}
\end{center} 
\end{figure}

For a given macro-ridge, increment in acceleration can be attained by reducing the length scale under which the pressure gradients are operative. When the pitch length $p$ is varied, the change to be incorporated into the model is to plug in an appropriate length scale. For $k_1$ substrate, the length scale does not change significantly. This can be inferred from the fact that the liquid encounters another ridge around the end of inertial regime. However, for $k_3$ substrate the pitch length $p$ is the proper length scale as the drop encounters multiple ridges during the inertial regime. With this notion, one can expect a higher acceleration. However, two adjacent ridges flush the liquid in the opposite direction at any valley. When $We > 25$, this even creates a Rayleigh jet which produces daughter droplets. Essentially, the lateral fragmentation of the drop is delayed. Furthermore, the liquid has to overcome the peak of a ridge during the recoiling at the periphery of the rim. Synergistically, these two effects degenerate the ability of the drop to lift-off early. The residence time in this case is considered as the time when the core of the drop bounces off from the surface if Rayleigh jet formation takes place. The non-dimensional residence time plot with $We$ is shown in Fig. \ref{contacttime_with_and_without_offset}a where $\tau_s$ represents the residence time on a substrate and it can be clearly seen that $k_3$ substrate has higher residence times in comparison with $k_0$ and $k_1$ substrates though it performs better compared to $k_\infty$ substrate. Surprisingly, the residence time on $k_2$ substrate in the regime $5 < We < 25$ is found to be even higher than that of $k_{\infty}$ substrate. For $We > 25$, the substrate follows similar trend to that of $k_0$ and $k_1$ because inertia allows the liquid ligaments to be detached from the surface under the influence of neighbor ridge (see supplementary videos 5 and 6). The detached ligaments are fed by the liquid mass recoiling at any ridge eliminating the flushing in the region between the ridges. Note that this behavior is observed even on $k_3$ substrate but the central film recoiling is greatly affected by multiple ridges as explained earlier. Increase in residence time is also predominant when the drop impinges the substrate at an offset to the ridge at low velocities ($We < 14$). Two offset distances are considered: $R_0/2$ and $R_0/4$ from the peak of ridge for $k_0$, $k_1$ and $k_2$ substrates. Again, in the case of $k_2$ substrate, the adjacent ridge greatly influences the lift-off as the residence time becomes $>$ 2.6$\tau$ for $We > 5$ which can be observed in Fig. \ref{contacttime_with_and_without_offset}b (see supplementary videos 11-14). Note that the details of $\tau_s/ \tau$ is not presented for $We > 14$ for $k_{2,2}$ and $k_{2,4}$ because of the daughter droplets formed by the jet which obstruct the light during side view imaging (see supplementary videos 15 and 16).

 Even though macro-texture deployment makes significant advancement in attaining rapid drop shedding, the design with multiple ridges involving another length scale $p$ is crucial as it might actually degrade the surface repellency. The lower limit of $p$ can be considered as $D_0$ since the performance with respect to residence time for both $k_0$ and $k_1$ substrates is found to be equally good. Any further reduction in $p$ will generate inferior lateral to vertical momentum conversion producing relatively slack bouncing behavior. Higher values of $p$ $(> D_0)$ will render a large overall area just equivalent to the flat superhydrophobic surface ($k_{\infty}$) and thus not leverage the effect of the macro-structure.

\subsection*{Methods}
\subsubsection*{Substrates}
The macro-ridges are machined on polished aluminium substrates (30x30x10 mm) with wirecut CNC electro disharge machining. The substrates are then coated with the superhydrophobic coating. The coating is a two step process involving bottom and top coats. The substrate is allowed to dry for one hour after the application of bottom coat and one day after the application of top coat under ambient atmospheric conditions. The bottom coat acts as a binder for the top coat. After complete drying, randomly rough micro-metric layer is formed on the substrate to render it superhydrophobic which can be observed in the SEM image of the $k_\infty$ substrate as shown in Fig. 2 of supplementary material.

\subsubsection*{Experimental Procedure}
The experimental setup consists of two high speed cameras (Photron SA4 and Photron Mini UX 100) synchronized together to capture the side and top views of the impact simultaneously. The side view resolution is 768 x 640 pixels and top view resolution is 1280 x 624 pixels. A Harvard syringe pump is used with a precision syringe and a calibrated needle with outer diameter of 0.72 mm to produce Milli-Q water drops of diameter, $D_0 =$ 2.95 $\mypm$ 0.03 mm. During the representation of substrates as $k_{n,m}$, the values of $n$ were found to be $n =$ 3$^{-0.02}_{-0.08}$ , 2$^{-0.01}_{-0.05}$ and 1 $^{-0.01}_{-0.03}$ while that of $m$ were found to be $m = $ 4$^{-0.14}_{-0.46}$ and 2 $^{-0.07}_{-0.33}$. A motorized stage is used to position the needle along the center of the ridge and also to control the drop release height with a spatial resolution of 2 $\mu$m. The side and top views are captured using diffused back light imaging and direct imaging techniques respectively. The frame rate is set at 8,000 fps with a shutter speed of 1/95,000 s and 1/81,920 s for the side and top views respectively. The diameter of drop and impact velocity are calculated from the images. All the images are processed using open source software ImageJ (1.50b). During centered impacts, the residence time is defined from the juncture when the nadir of drop contacts the substrate to the point in time when the fragmented parts completely lift-off except for $k_3$ substrate where the core drop lift-off is considered. During impingement at offsets, when any portion of surface of the drop contacts the substrate, it is considered to be the origin of residence time. For each height the experiment is conducted three times to ensure the repeatability of the phenomenon.


\subsection*{Acknowledgements}
We thank NFMTC of Indian Institute of Technology, Madras for providing the SEM imaging facility.

\subsection*{Author contributions}
R.K. and S.B. designed the experiments. R.K. performed the experiments. R.K., S.B. and S.K.D. discussed the model. R.K., S.B. and S.K.D. wrote the paper.

\subsection*{Additional Information}
\textbf{Supplementary Information} accompanies this paper.\\
\textbf{Competing financial interests} The authors declare no competing financial interests.

\end{document}